\documentclass[doublecol,figures,letterpaper]{epl2} 
\usepackage{amsmath}
\usepackage{amssymb}

\newcommand{\mb}[1]{\mathbf{#1}}
\newcommand{\bs}[1]{\boldsymbol{#1}}

\title{Global cross-over dynamics of single semiflexible polymers}

\author{M. Hinczewski \and R.R. Netz}

\institute{                    
  Department of Physics, Technical University of
Munich - 85748 Garching, Germany
}


\abstract{We present a mean-field dynamical theory for single
  semiflexible polymers which can precisely capture, without fitting
  parameters, recent fluorescence correlation spectroscopy results on
  single monomer kinetics of DNA strands in solution. Our approach
  works globally, covering three decades of strand length and five
  decades of time: it includes the complex cross-overs occurring
  between stiffness-dominated and flexible bending modes, along with
  larger-scale rotational and center-of-mass motion.  The accuracy of
  the theory stems in part from long-range hydrodynamic coupling
  between the monomers, which makes a mean-field description more
  realistic.  Its validity extends even to short, stiff fragments,
  where we also test the theory through Brownian hydrodynamics
  simulations.}
\begin{document}

\maketitle

For a polymer in solution, hydrodynamics introduces long-range
coupling between different points on the chain contour, its strength
falling off with inverse distance like $1/r$.  Though this has long
been recognized as a crucial factor in understanding the dynamics of
flexible polymers~\cite{DoiEdwards}, its importance in the case of
semiflexible and stiff chains is not fully appreciated.  The cause is
often expediency: even without taking long-range coupling into
account, the nature of the local interactions---governed by bending
stiffness and inextensibility---presents a formidable problem in
constructing a theory of semiflexible polymer dynamics.  The most
widely used model---the worm-like chain
(WLC)~\cite{Kratky1949}---yields nonlinear equations of motion, and
thus all dynamical theories of the WLC have been approximate.  One
approach is the weakly-bending assumption: deriving equations of
motion from a perturbation analysis around the rigid rod
limit~\cite{Granek1997,Kroy1997,Gittes1998,Everaers1999,Hallatschek2005}.
This is particularly relevant for certain biopolymers like actin,
where the large persistence length, $l_p \sim {\cal O}(1\:
\mu\text{m})$, means that a broad dynamical regime, consisting of
motion on length scales much smaller than $l_p$, will be dominated by
the bending stiffness.  Yet for less rigid cases like double-stranded
DNA, where $l_p \approx 50$ nm, many empirical situations will involve
complex cross-overs between stiffness-dominated and flexible regimes
at different time scales.  Weakly-bending approaches cannot provide an
accurate description of these cross-overs, which hydrodynamics makes
even more challenging to model: the fluctuation modes at all length
scales are coupled due to the long-range interactions.

The need for a comprehensive, quantitatively accurate theory of
semiflexible polymer dynamics is made more urgent by advances in
single-molecule experimental techniques. Fluorescence correlation
spectroscopy (FCS) can already probe double-stranded DNA kinetics at
the level of a single
monomer~\cite{Lumma2003,Shusterman2004,Petrov2006}.  The most recent
results, by Petrov {\it et. al.}~\cite{Petrov2006}, reveal a rich
sequence of dynamical behaviors for the motion of a tagged end in DNA
fragments varying in length from $L \approx 30-7000$ nm.  In our work,
we show that a single theory, without any fitting parameters, can give
an excellent description of these experimental results over the entire
range, covering the whole cross-over between stiff and flexible chain
dynamics.  Our approach is based on a Gaussian mean-field theory
(MFT)~\cite{Winkler1994,Ha1995,Harnau1996,Winkler2007,Hinczewski2009},
which is in itself surprising: Gaussian models are usually considered
tools for flexible polymers, with limited applicability as one nears
the rigid rod limit.  What we demonstrate is that hydrodynamics---the
complicating element in the theory---is what underlies its success:
the MFT becomes more accurate because of the long-range interactions.
We specifically concentrate on highly stiff chains, the most difficult
terrain for a Gaussian theory, and investigate the strengths and
limitations of the MFT approximation.  Since the experimental results
for short-time dynamics of stiff fragments are not well resolved, we
additionally validate the theory through Brownian hydrodynamics (BD)
simulations.  Though the MFT is restricted to quantities which are
spatially-averaged over all coordinate directions (i.e.~the mean
square displacements measured through FCS), it can capture the
full complexity of the monomer motion: hydrodynamic effects,
rotational and center-of-mass diffusion, and the cross-overs between
dynamical regimes at different time scales.

We begin by reviewing the general MFT approach to semiflexible polymer
dynamics (a more detailed treatment can be found in
Ref.~\cite{Hinczewski2009}).  Polymer stiffness is typically modeled
by the WLC Hamiltonian, $U_\text{WLC} = \frac{1}{2}l_pk_B T \int ds\,
(\partial_s \mb{u}(s))^2$, describing the bending energy of a polymer
contour $\mb{r}(s)$, $0 \le s \le L$, with persistence length $l_p$
and tangent vector $\mb{u}(s) \equiv \partial_s \mb{r}(s)$.  The local
inextensibility of the contour is expressed through the restriction
$|\mb{u}(s)|=1$ at each $s$.  The partition function $Z$ is a path
integral over all possible contours, $Z = \int {\cal D}\mb{u}\,
\prod_s \delta(|\mb{u}(s)|-1)\,\exp(-\beta U_\text{WLC})$, with the
$\delta$ functions enforcing the inextensibility constraint and $\beta
= 1/(k_B T)$.  Since this constraint is nonlinear, only certain
equilibrium properties can be calculated exactly, and deriving any
dynamical quantities requires an approximation.  In our case, we
obtain a more tractable Gaussian mean-field model by estimating $Z$
through the stationary phase approach \cite{Ha1995}, $Z \approx
\exp(-\beta {\cal F}_\text{MF}) = \int {\cal D}\mb{u}\, \exp(-\beta
U_\text{MF})$, where the MFT Hamiltonian $U_\text{MF}$ has the form:
$U_\text{MF}=(\epsilon/2) \int ds\,\left(\partial_s \mb{u}(s)\right)^2
+\nu \int ds\,\mb{u}^2(s) + \nu_0 \left(\mb{u}^2(0) +
\mb{u}^2(L)\right)$.  Here local inextensibility has been relaxed, and
the parameters $\nu$ and $\nu_0$ are related through the stationary
phase condition, $\partial_{\nu} {\cal F}_\text{MF} = \partial_{\nu_0}
{\cal F}_\text{MF} = 0$.  The latter yields $\sqrt{\nu \epsilon/2} =
\nu_0 = 3k_BT/4$.  This is equivalent to making $\nu$ and $\nu_0$ act
as Lagrange multipliers enforcing the global and end-point constraints
$\int ds\, \langle \mb{u}^2(s)\rangle = L$, $\langle \mb{u}^2(0)
\rangle = \langle \mb{u}^2(L)\rangle = 1$.  By setting the bending
modulus $\epsilon = (3/2)l_p k_BT$, the Hamiltonian $U_\text{MF}$ can
be tuned to reproduce exactly the tangent-tangent correlation $\langle
\mb{u}(s) \cdot \mb{u}(s^\prime) \rangle = \exp(-|s-s^\prime|/l_p)$ of
a WLC with persistence length $l_p$, as well as related thermodynamic
averages like the mean square end-to-end vector $\langle \mb{R}^2
\rangle$.

The Gaussian approximation provides a starting point
for deriving the dynamics of the system, following a hydrodynamic
pre-averaging approach similar to that of the Zimm
model~\cite{Harnau1996,Hinczewski2009}.  The behavior of the
chain contour $\mb{r}(s,t)$ obeys the Langevin equation, $\partial_t
\mb{r}(s,t) = -\int ds^\prime\, \mu_\text{avg}(s-s^\prime)\delta
U_\text{MF}/\delta \mb{r}(s^\prime,t) + \bs{\xi}(s,t)$, where
$\bs{\xi}(s,t)$ are Gaussian stochastic velocities whose components
have correlations governed by the fluctuation-dissipation theorem:
$\langle \xi_i(s,t) \xi_j(s^\prime,t^\prime)\rangle = 2 k_B T
\delta_{ij}\delta(t-t^\prime)\mu_\text{avg}(s-s^\prime)$.  Here $\mu_\text{avg}(s-s^\prime)$ is the pre-averaged
mobility tensor, obtained from the continuum Rotne-Prager tensor
$\overleftrightarrow{\bs{\mu}}(s,s^\prime;\mb{x})$ \cite{Rotne1969,
  Harnau1996} describing long-range hydrodynamic interactions between
two points $s$, $s^\prime$ on the contour at spatial separation
$\mb{x}$:
\begin{equation}\label{t7}
\begin{split}
&\overleftrightarrow{\bs{\mu}}(s,s^\prime;\mb{x}) = 2a\mu_0 \delta(s-s^\prime)\overleftrightarrow{\mb{1}}+ \Theta(x-2a)\\
&\quad \cdot \left(\frac{1}{8\pi\eta x} \left[\overleftrightarrow{\mb{1}} +\frac{\mb{x}\otimes\mb{x}}{x^2}\right] + \frac{a^2}{4\pi\eta x^3}\left[ \frac{\overleftrightarrow{\mb{1}}}{3} - \frac{\mb{x}\otimes\mb{x}}{x^2}\right]\right)\,.
\end{split}
\end{equation}
$\overleftrightarrow{\mb{1}}$ is the $3\times 3$ identity matrix, $a$
is a microscopic length scale corresponding to the monomer radius,
$\eta$ is the viscosity of water, $\mu_0 = 1/6\pi \eta a$ is the
Stokes mobility of a sphere of radius $a$, and the $\Theta$ step
function excludes unphysical situations involving overlap between
monomers.  If a polymer configuration with points $s$ and $s^\prime$
separated by $\mb{x}$ has an equilibrium probability
$G(s,s^\prime;\mb{x})$, then the pre-averaged mobility is defined
through the integration: $\int
d^3\mb{x}\,\overleftrightarrow{\bs{\mu}}(s,s^\prime;\mb{x})
G(s,s^\prime;\mb{x}) = \mu_\text{avg}(s-s^\prime)
\overleftrightarrow{\mb{1}}$.  In order to determine the relative
importance of hydrodynamic effects, we will also compare the
free-draining case where the long-range interactions are turned off,
and the mobility reads $\mu^\text{fd}_\text{avg}(s-s^\prime) = 2a\mu_0
\delta(s-s^\prime)$.

The Langevin equation can be solved through normal mode decomposition,
yielding a set of coupled stochastic partial differential equations
for the normal mode amplitudes.  These can be diagonalized, resulting
in a set of decoupled normal modes characterized by relaxation times
$\tau_n$ (in decreasing order such that $\tau_1$ is largest).  As in
Ref.~\cite{Hinczewski2009}, we impose a mode number cutoff $M = L/8a$
to agree with BD simulations at short times, approximately modeling
the discrete nature of the chain at distances on the order of the
monomer radius $a$.  Results at larger length scales (i.e. the
experimental comparisons discussed below) do not depend on the details
of the cutoff.

To test the accuracy of the MFT, we compared the theoretical results
to BD simulations~\cite{Ermak1978} of a bead-spring worm-like chain
consisting of $N$ monomers of radius $a$ (contour length $L=2aN$).
The bead positions $\mb{r}_i(t)$ obey a discrete Langevin equation,
with monomers coupled hydrodynamically through the Rotne-Prager
tensor~\cite{Rotne1969}.  The elastic potential of the chain $U =
U_\text{ben} + U_\text{str} + U_\text{LJ}$ has three parts: (i) a
bending energy $U_\text{ben} = (l_p k_B T/2a)\sum_{i}
(1-\cos\theta_i)$, where $\theta_i$ is the angle between adjacent
bonds; (ii) a harmonic stretching term $U_\text{str} = (\gamma/4a)
\sum_i \left(r_{i+1,i}-2a\right)^2$, where the large modulus $\gamma =
200 k_B T /a$ ensures inextensibility and $\mb{r}_{i,j}
=\mb{r}_i-\mb{r}_j$; (iii) a truncated Lennard-Jones interaction
$U_\text{LJ} = \omega \sum_{i < j}
\Theta(2a-r_{i,j})[(2a/r_{i,j})^{12}-2 (2a/r_{i,j})^{6} +1]$ with
$\omega = 3k_B T$.  In our simulations we set $N=50$, with various
$l_p= 10a-200a$, and used a Langevin time step $\tau = 3 \times
10^{-4}\: a^2/(k_BT\mu_0)$.  Thermodynamic averages were derived from
5-20 independent runs, with data collection occurring every
$10^2-10^3$ steps, and each run lasting $10^8 - 10^9$ steps.

\begin{figure}[t]
\onefigure[width=\columnwidth]{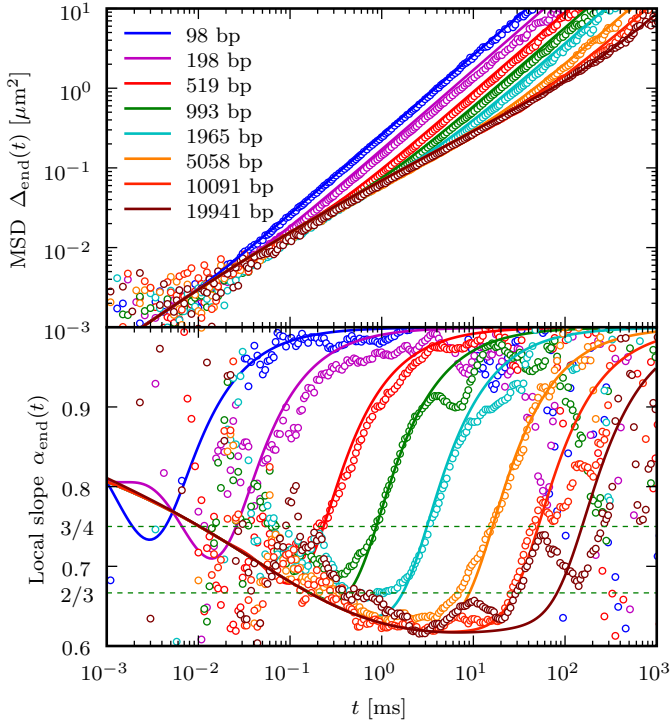}
\caption{Top: $\Delta_\text{end}(t)$, the mean square displacement of
  an end-monomer in a dsDNA strand, for various lengths $L = 98-19941$
  bp.  Bottom: the local slope $\alpha_\text{end}(t) = d\ln
  \Delta_\text{end}/d\ln t$ of the log-log curves in the top panel.
  In both panels the circles are from the experimental FCS
  measurements in Ref.~\cite{Petrov2006}.  The solid lines are the MFT
  predictions, without any fitting parameters: all values are taken
  from the literature and experimental conditions (see text).}
\label{petrov}
\end{figure}

We start our analysis by focusing on one specific dynamical quantity:
the mean square displacement (MSD) of a chain end-point,
$\Delta_\text{end}(t) \equiv \langle (\mb{r}(L,t) - \mb{r}(L,0))^2
\rangle$.  The top panel of Fig.~\ref{petrov} shows
$\Delta_\text{end}(t)$ for DNA strands of various length, $L \approx
100 - 20000$ bp, taken from the FCS measurements of Petrov {\it
  et. al.}~\cite{Petrov2006}.  Superimposed is the MFT prediction,
without any fitting parameters.  The constants in the theory are taken
from the experimental conditions and the literature: $T = 298$ K,
$\eta = 0.891$ $\text{mPa}\cdot\text{s}$, $a = 1$ nm, a rise per bp of
$0.34$ nm, $l_p = 50$ nm.  The agreement between the MFT and
experimental results is remarkable: the average discrepancy in the
time range $t = 10^{-1}-10^2$ ms, where there is the least scatter in
the FCS data, varies between $6-25\%$ for the different $L$.  This
close agreement without fitting parameters is only possible if the
full set of equations for the MFT normal mode amplitudes, including
the off-diagonal coupling between normal modes due to hydrodynamics,
is diagonalized and solved.  If the off-diagonal elements are assumed
negligible, as was done in Ref.~\cite{Petrov2006}, additional fitting
parameters are required to get agreement: rescaling factors for the
relaxation times and the diffusion constant.  This in itself is a
testament to the importance of hydrodynamic effects.  In
Ref.~\cite{Hinczewski2009}, we had compared the MFT to FCS data from a
similar experiment~\cite{Shusterman2004} for a smaller set of DNA
chain lengths: the discrepancies we observed between MFT and
experiment at intermediate times are completely absent in the more
recent and extensive FCS results analyzed here.  This might indicate
that the experimental setup or analysis in Ref.~\cite{Shusterman2004}
may need to be re-examined.

The local slope of the $\Delta_\text{end}(t)$
curves in the log-log plot, $\alpha_\text{end}(t)$, is shown in the
bottom panel of Fig~\ref{petrov}.  $\alpha_\text{end}(t)$ is
calculated for each time $t$ by fitting a straight line to the log-log
plot of data points within a small range of times $t_i$ defined by
$|\log_{10} t_i/t| < 0.15$.  The local slope would be constant for
pure power law behavior, and what we find as $L$ is increased is the
gradual emergence of a scaling regime at intermediate times with
$\alpha_\text{end} \approx 0.62$.  The standard expectation for long,
flexible polymers in solution is given by the Zimm result,
$\alpha_\text{end} = 2/3$~\cite{DoiEdwards}.  In our case, the sub-Zimm
scaling and pronounced variation in $\alpha_\text{end}(t)$ with $t$ is
evidence that slow cross-over effects have a significant
role~\cite{Hinczewski2009}: the resulting deviation from the classical
scaling theory of flexible polymers is thus experimentally observable,
and precisely captured by the MFT.

The time at which the slope reaches its minimum value is the same
order of magnitude as $\tau_1$, the longest relaxation time.  For $t >
\tau_1$ we see a cross-over to center-of-mass motion, with a slope
$\alpha_\text{end} = 1$.  The oscillations in the FCS slopes in this
cross-over are due to uncertainties in $\Delta_\text{end}(t)$ at large
$t$, arising from noise in the long-time exponential tails of the FCS
correlation functions.  There should be another cross-over at short
times, where the length scale of the fluctuations is comparable to the
persistence length, $\Delta_\text{end}(t) \lesssim l_p^2 \approx
2.5\times 10^{-3}$ $\mu$m$^2$.  Here the expected behavior is a
power-law scaling with exponent $\alpha_\text{end}=3/4$ in the
free-draining case; hydrodynamics introduces logarithmic corrections
to this scaling, which are observable as an increase in the exponent
on the order of $10\%$~\cite{Granek1997,Gittes1998}.  Unfortunately in
the time range where we should see this stiffness-dominated regime the
FCS data is not well-resolved: there is too much scatter in the
$\Delta_\text{end}(t)$ results when $t < 10^{-1}$ ms for us to be able
to extract accurate local slopes. Note that the MFT curves are still
in excellent agreement even for the shortest chains examined, where
$L$ is smaller or comparable to $l_p \approx 50$ nm ($\approx 150$
bp).  But what is being measured for these short, stiff fragments is
essentially only the cross-over to center-of-mass diffusion, with the
slope $\alpha_\text{end}$ approaching 1.

\begin{figure}[t]
\onefigure[width=\columnwidth]{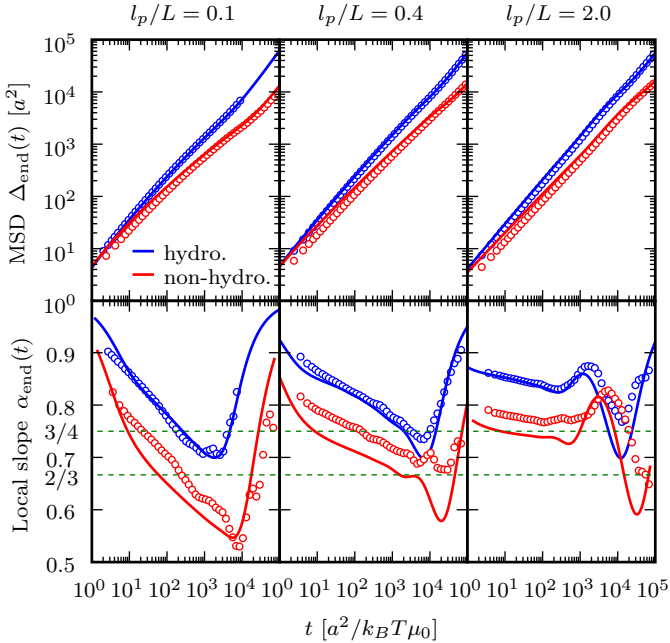}
\caption{Top: $\Delta_\text{end}(t)$, the MSD of
  an end-monomer in a semiflexible polymer, for $L = 100a$ and three
  different ratios $l_p/L$ ($a$ = monomer radius).  Bottom: the local
  slope $\alpha_\text{end}(t) = d\ln \Delta_\text{end}/d\ln t$ of the
  log-log curves in the top panels.  In all panels solid lines are MFT
  predictions, while circles are BD simulation results.  The upper
  curves (blue) include long-range hydrodynamic interactions, in
  contrast to the lower curves (red) which correspond to the
  free-draining limit.}
\label{hydro}
\end{figure}

We thus need an alternative approach to validate the MFT description
of polymer dynamics in the cross-over to the stiff regime.  In this
case BD simulations are an ideal tool: computational constraints
restrict us to relatively short chains, but this is precisely the
limit we want to investigate, where $l_p \gg a$, $L \sim
\text{O}(l_p)$.  Fig.~\ref{hydro} shows the comparison between
simulation and MFT $\Delta_\text{end}(t)$ for a chain of length
$L=100a$ and varying $l_p$, going from the flexible case of $l_p/L
=0.1$, to the stiff one of $l_p/L = 2.0$.  The regular
theory/simulation results with hydrodynamic interactions (shown in
blue) are contrasted to the free-draining case (shown in red).  We see
a quite interesting trend: the hydrodynamic MFT continues to be an
accurate predictor of $\Delta_\text{end}(t)$, even as $l_p$ becomes
larger than $L$.  In the time range $t = 10^{1} - 10^{4}$ $a^2/k_BT
\mu_0$ the average error between the hydrodynamic MFT and simulation
$\Delta_\text{end}(t)$ varies from $3-15\%$ for the different $l_p$,
similar to the errors seen in the experimental comparison above.  On
the other hand, the free-draining MFT is much less accurate: it
noticeably overestimates the short-time $\Delta_\text{end}(t)$, and
the average errors compared to the BD results are 4-8 times larger
than in the hydrodynamic counterparts.  This is also plainly seen in
the local slopes plotted in the bottom panels of Fig.~\ref{hydro}: the
non-hydrodynamic MFT performs significantly worse.  Though
hydrodynamics introduces another level of complexity into our
approach, requiring an additional approximation in the form of
pre-averaging, the resulting MFT is quantitatively more successful
than the simpler free-draining theory.  This reflects a general
well-known feature of mean-field theories: they are closer to reality
in systems with long-range interactions.  Due to hydrodynamics, every
point on the chain is coupled to every other point through the
Rotne-Prager tensor. For a free-draining polymer, the dynamics of a
point on the chain are determined solely by the local
bending/extensibility interactions with its nearest neighbors.  Thus
we can expect a Gaussian mean-field description in this case to be a
much cruder estimate.

For the BD simulation results in Fig.~\ref{hydro}, the evolution of
the dynamics from the flexible to stiff limits roughly agrees with
earlier scaling theory expectations: for the free-draining chain with
$l_p/L = 0.1$ we see an intermediate time regime with
$\alpha_\text{end}$ approaching the Rouse value of 1/2, appropriate
for flexible polymers~\cite{DoiEdwards}, which increases to around 3/4
as we move to larger $l_p$.  Adding hydrodynamics has several effects:
the center-of-mass diffusion becomes faster (due to hydrodynamic
entrainment), the relaxation times become shorter, and the local
slopes are all shifted upwards.  The $\alpha_\text{end}$ plateau near
3/4 in the stiff limit is now near $0.85$.  For both the hydrodynamic
and free-draining cases, note that only at the largest $l_p$ do we see
the emergence of nearly pure power-law scaling behavior at short
times: cross-over effects dominate in the more flexible cases.
Moreover the shape of the $\alpha_\text{end}(t)$ curve changes as
$l_p$ is increased: two local minima are formed, a shallow one in the
short-time plateau region, and a much sharper dip at longer times on
the order of $\tau_1$.  This sharp dip, connected to the appearance of
a new dynamical regime, will be discussed further below.  First we
turn to a more basic question: given that at $l_p/L = 2.0$ we are
approaching the stiff rod limit, how is it that a Gaussian model can
still work so well?

\begin{figure}[t]
\includegraphics*[width=\columnwidth]{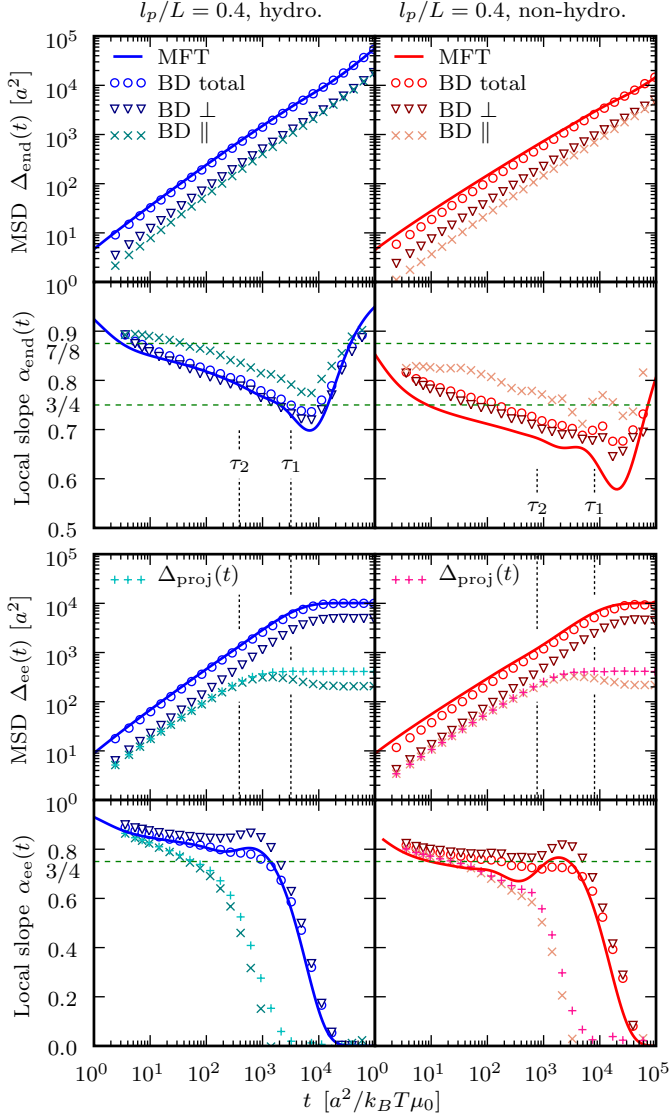}
\caption{Top panels: the end-point MSD $\Delta_\text{end}(t)$ and
  local slopes $\alpha_\text{end}(t)$ for a chain with $L=100a$ and
  $l_p/L =0.4$.  Both hydrodynamic and free-draining results are
  shown.  MFT estimates for the total $\Delta_\text{end}(t)$ are drawn
  as solid lines.  The BD simulation results show both the total
  $\Delta_\text{end}(t)$ and the decomposition into components,
  $\Delta^\parallel_\text{end}(t)$, $\Delta^\perp_\text{end}(t)$.
  Vertical lines mark the two largest relaxation times, $\tau_1$ and
  $\tau_2$, calculated from the MFT. Bottom panels: analogous results
  for the end-to-end MSD $\Delta_\text{ee}(t)$ and corresponding local
  slope $\alpha_\text{ee}(t)$.  The projected length MSD,
  $\Delta_\text{proj} = \langle (R(t)-R(0))^2 \rangle$, is plotted as
  well for comparison.}
\label{lp40}
\end{figure}

\begin{figure}[t]
\includegraphics*[width=\columnwidth]{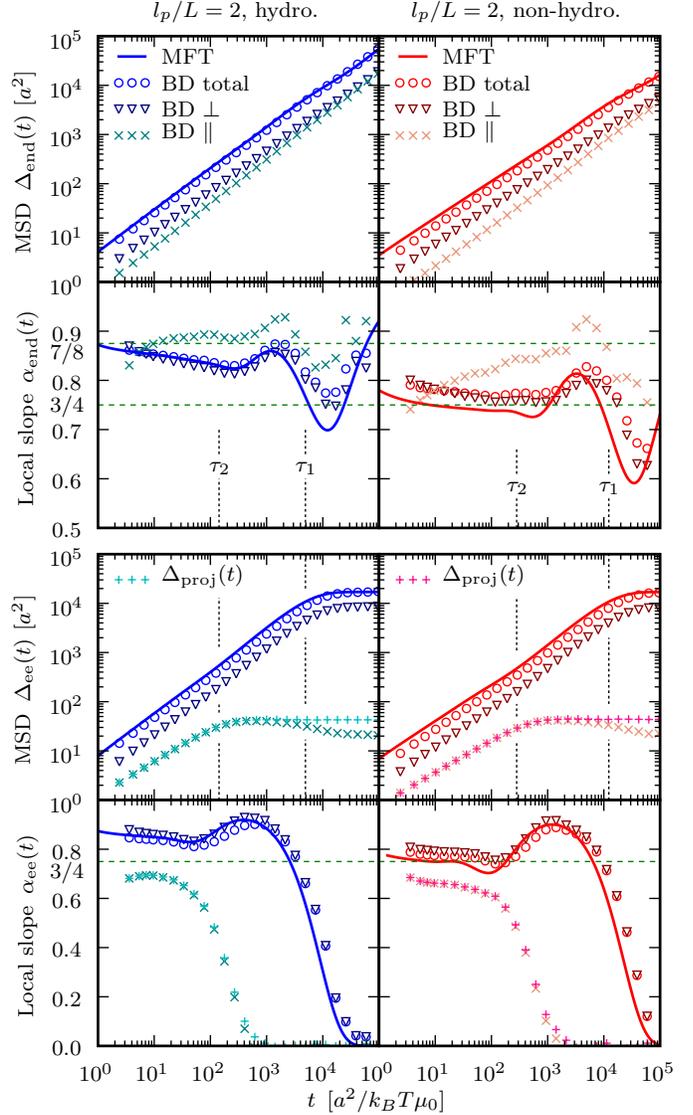}
\caption{Same as in Fig.~\ref{lp40}, but for $l_p/L = 2.0$}
\label{lp200}
\end{figure}

\begin{figure}[t]
\centering\includegraphics*[width=\columnwidth]{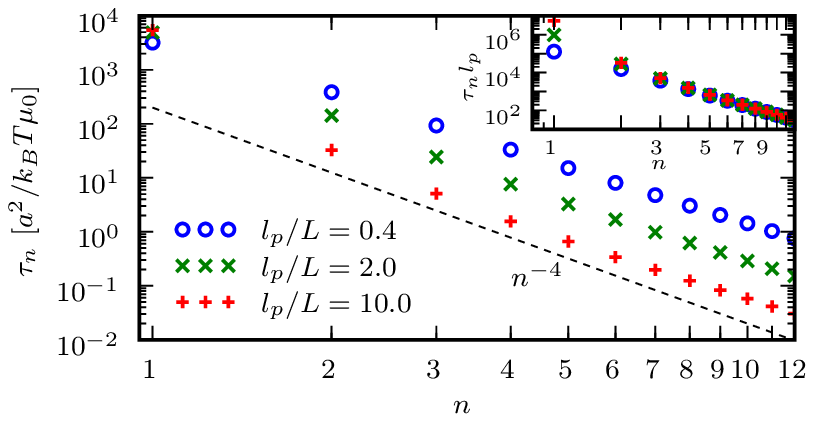}
\caption{Relaxation times $\tau_n$ versus mode number $n$ calculated
  using the MFT for a semiflexible polymer with $L=100a$ and various
  $l_p/L$.  The dashed line shows an $n^{-4}$ power-law scaling.  Inset: same data, but plotted as $\tau_n l_p$ versus $n$.}
\label{tau}
\end{figure}

To answer this question, we look at the dynamics of semiflexible and
stiff chains in more detail, by decomposing the motion into components
approximately parallel and perpendicular to the chain contour.  For
the chain end-point, we define a tangent vector
$\hat{\mb{u}}_\text{avg}(t) \equiv (\mb{u}(L,t)+
\mb{u}(L,0))/|\mb{u}(L,t)+ \mb{u}(L,0)|$, which is an average between
the end-point tangent at times $t$ and $0$.  The parallel and
perpendicular components of $\Delta_\text{end}(t)$ are then given by:
$\Delta^\parallel_\text{end}(t) = \langle
\left[(\mb{r}(L,t)-\mb{r}(L,0))\cdot
  \hat{\mb{u}}_\text{avg}(t)\right]^2 \rangle$,
$\Delta^\perp_\text{end}(t) = (\Delta_\text{end}(t) -
\Delta^\parallel_\text{end}(t))/2$.  This decomposition is applied to
BD simulation results of a chain with $L=100a$ and $l_p/L = 0.4$,
$2.0$ in the top panels of Figs.~\ref{lp40}-\ref{lp200}.  The total
MSD, $\Delta_\text{end}(t)$, and the corresponding MFT estimate is
also plotted for comparison, and all quantities are shown both in the
hydrodynamic and free-draining cases.  At times greater than the
largest relaxation time, $t \gtrsim \tau_1$, where the orientation of
$\mb{u}(L,t)$ and $\mb{u}(L,0)$ are uncorrelated, the components
converge to the same value, equal to 1/3 the total
$\Delta_\text{end}(t)$.  For shorter times we see an anisotropy,
becoming more prominent as the chain becomes stiffer:
$\Delta^\parallel_\text{end}(t) < \Delta^\perp_\text{end}(t)$, due to
the suppression of parallel fluctuations with increasing $l_p$.  In
the absence of hydrodynamics, one can derive scaling predictions for
the two components based on the weakly-bending approximation:
$\Delta^\perp_\text{end}(t)$ should have an exponent
$\alpha^\perp_\text{end} = 3/4$~\cite{Granek1997,Gittes1998}, while
the effects of tension propagation along the backbone lead to a higher
exponent $\alpha^\parallel_\text{end}=7/8$ for
$\Delta^\parallel_\text{end}(t)$~\cite{Everaers1999,Hallatschek2005}.
We see these two dynamical behaviors most clearly in the free-draining
BD results for $l_p/L = 2.0$, with the local slopes
$\alpha_\text{end}^\perp(t)$ and $\alpha_\text{end}^\parallel(t)$
approaching 3/4 and 7/8 at intermediate times (with slight
modifications due to cross-over effects).  Adding hydrodynamics
changes the scaling, shifting both of these exponents up by $5-15\%$.

The MFT entirely misses the anisotropy: given the isotropic nature of
the Gaussian Hamiltonian, it predicts $\Delta^\parallel_\text{end}(t)
= \Delta^\perp_\text{end}(t)$.  However, the MFT will still work for
any quantity which is averaged over all spatial directions, and thus
can successfully estimate the total $\Delta_\text{end}(t)$.  In the
stiff limit, the spatially averaged MSD is dominated by large
fluctuations perpendicular to the backbone, for which the Gaussian MFT
gives a reasonable description; in fact, as we will argue below, the
MFT normal modes at large $l_p$ for $n>1$ effectively behave like the
transverse modes in the standard weakly-bending perturbation analysis.
In contrast, the longitudinal fluctuations of a nearly stiff rod
cannot be approximated well by a Gaussian model.  But because their
contribution to the total $\Delta_\text{end}(t)$ is small, the MFT
manages to capture the spatially-averaged dynamics.  While these
limitations are not relevant to modeling the FCS results for
$\Delta_\text{end}(t)$ of a freely diffusing chain, they would be
significant for an anisotropic experimental setup, i.e. a chain under
tension. In this case the MFT would have to be
refined~\cite{Hinczewski2009b}.

Another way of looking at the chain internal kinetics is through the
end-to-end vector MSD, $\Delta_\text{ee}(t) \equiv \langle
(\mb{R}(t)-\mb{R}(0))^2 \rangle$, where $\mb{R}(t) = \mb{r}(L,t) -
\mb{r}(0,t)$.  For the same parameters used in the top panels of
Figs.~\ref{lp40}-\ref{lp200}, the analogous $\Delta_\text{ee}(t)$ results are shown
in the bottom panels, together with the corresponding local slopes
$\alpha_\text{ee}(t)$.  The decomposition here is defined with respect
to the unit vector $\hat{\mb{R}}_\text{avg}(t) \equiv (\mb{R}(t)+
\mb{R}(0))/|\mb{R}(t)+ \mb{R}(0)|$, so that
$\Delta^\parallel_\text{ee}(t) = \langle [(\mb{R}(t)-\mb{R}(0))\cdot
  \hat{\mb{R}}_\text{avg}(t)]^2 \rangle$, $\Delta^\perp_\text{ee}(t) =
(\Delta_\text{ee}(t) - \Delta^\parallel_\text{ee}(t))/2$.  The
component $\Delta^\parallel_\text{ee}(t)$ at short times is
approximately equal to the MSD of the projected length,
$\Delta_\text{proj}(t) = \langle (R(t) - R(0))^2 \rangle$, where $R(t)
= |\mb{R}(t)|$.  This is a well-studied quantity for characterizing
semiflexible polymer dynamics~\cite{Granek1997,LeGoff2002}, and is
also plotted in Figs.~\ref{lp40}-\ref{lp200}. In the stiff limit, the time scale at
which $\Delta_\text{proj}(t)$ saturates is the upper bound for
relaxation of both $\parallel$ and $\perp$ internal contour
fluctuations. As can be seen in Figs.~\ref{lp40}-\ref{lp200}, this saturation time
coincides approximately with $\tau_2$.  Though
$\Delta^\perp_\text{ee}(t)$ continues to grow between $\tau_2$ and
$\tau_1$, the dominant contribution in this range is rotational
diffusion of the polymer backbone.  The underlying reason is that for
$l_p/L \to \infty$, the $n=1$ normal mode in the MFT becomes a purely
rotational mode~\cite{Harnau1995,Winkler2007}.  We can see this
directly in the plot of $\tau_n$ versus $n$ for $L=100a$ and various
$l_p/L$ in Fig.~\ref{tau}: as we move to stiffer chains, $\tau_1$
approaches a constant value, independent of $l_p$.  Up to a prefactor,
this constant agrees with $\tau_r \sim L^3/\mu_0 k_BT \ln(L/2a)$, the
rotational relaxation time of a stiff rod of length $L$ and diameter
$2a$.  The behavior of $\tau_n$ for $n>1$ is quite different: the
times scale approximately like $\sim L^4/l_p n^{4}$ (the inset of
Fig.~\ref{tau} shows data collapse for different $l_p$), with
corrections due to hydrodynamics.  This is exactly the predicted
behavior of the relaxation times for transverse fluctuation modes in
the weakly bending approach~\cite{Granek1997}.

Thus we can now understand fully the various dynamical regimes seen in
both the simulation and MFT results for $\Delta_\text{end}(t)$ in the
stiff limit.  For short times $t \lesssim \tau_2$ the motion is
dominated by the $\perp$ bending modes of the chain, and we see the
characteristic $3/4$ scaling (plus hydrodynamic corrections) in the
total MSD curves.  Between $\tau_2$ and $\tau_1$ we have a regime
controlled by rotational diffusion: the local slopes increase after
$\tau_2$, dip sharply through $\tau_1$, before rising to 1 at times $t
\gg \tau_1$, where center-of-mass translational diffusion is dominant.
This rotational regime has not been correctly reproduced by any
earlier theory based on the weakly-bending assumption.  Though the MFT
is limited to spatially-averaged properties, it does describe
quantitatively the full cross-over between all three regimes,
particularly important when making detailed comparisons to
experiments.

In summary, we have presented a mean-field approach to semiflexible
polymer dynamics that gives a highly accurate description, without
fitting parameters, of both experimental FCS measurements on DNA and
BD simulations.  It incorporates hydrodynamics and works over a wide
range of flexibility, even for short, stiff fragments.  The latter
case is particularly interesting with respect to DNA: there are claims
that the elastic energy of the WLC model may no longer be applicable
at length scales $\lesssim 100$ nm~\cite{Wiggins2006}.  If the
resolution of FCS experiments could be increased to probe this regime,
it would provide a direct and independent test, together with our
theory and simulation results, of DNA mechanical properties at these
scales, highly relevant to cellular processes.

\acknowledgments We thank R. Winkler and E. Petrov for useful
discussions, the Gilgamesh Cluster at the Feza G\"ursey Institute for
computing resources, and the Excellence Cluster ``Nano-Initiative
Munich'' for financial support.

\bibliographystyle{eplbib}
\bibliography{polyref_stiff}

\begin{thebibliography}{10}
\expandafter\ifx\csname url\endcsname\relax\def\url#1{\texttt{#1}}\fi

\bibitem{DoiEdwards}
\Name{Doi M. \and Edwards S.~F.} \Book{The Theory of Polymer Dynamics} ({Oxford
  University Press}) 1988.

\bibitem{Kratky1949}
\Name{Kratky O. \and Porod G.} \REVIEW{Rec. Trav. Chim. Pays-Bas
  }{68}{1949}{1106}.

\bibitem{Granek1997}
\Name{Granek R.} \REVIEW{J. Phys. II (France) }{7}{1997}{1761}.

\bibitem{Kroy1997}
\Name{Kroy K. \and Frey E.} \REVIEW{Phys. Rev. E }{55}{1997}{3092}.

\bibitem{Gittes1998}
\Name{Gittes F. \and MacKintosh F.~C.} \REVIEW{Phys. Rev. E }{58}{1998}{R1241}.

\bibitem{Everaers1999}
\Name{Everaers R., J\"ulicher F., Ajdari A. \and Maggs A.~C.} \REVIEW{Phys.
  Rev. Lett. }{82}{1999}{3717}.

\bibitem{Hallatschek2005}
\Name{Hallatschek O., Frey E. \and Kroy K.} \REVIEW{Phys. Rev. Lett.
  }{94}{2005}{077804}.

\bibitem{Lumma2003}
\Name{Lumma D., Keller S., Vilgis T. \and R\"adler J.~O.} \REVIEW{Phys. Rev.
  Lett. }{90}{2003}{218301}.

\bibitem{Shusterman2004}
\Name{Shusterman R., Alon S., Gavrinyov T. \and Krichevsky O.} \REVIEW{Phys.
  Rev. Lett. }{92}{2004}{048303}.

\bibitem{Petrov2006}
\Name{Petrov E.~P., Ohrt T., Winkler R.~G. \and Schwille P.} \REVIEW{Phys. Rev.
  Lett. }{97}{2006}{258101}.

\bibitem{Winkler1994}
\Name{Winkler R.~G., Reineker P. \and Harnau L.} \REVIEW{J. Chem. Phys.
  }{101}{1994}{8119}.

\bibitem{Ha1995}
\Name{Ha B.~Y. \and Thirumalai D.} \REVIEW{J. Chem. Phys. }{103}{1995}{9408}.

\bibitem{Harnau1996}
\Name{Harnau L., Winkler R.~G. \and Reineker P.} \REVIEW{J. Chem. Phys.
  }{104}{1996}{6355}.

\bibitem{Winkler2007}
\Name{Winkler R.~G.} \REVIEW{J. Chem. Phys. }{127}{2007}{054904}.

\bibitem{Hinczewski2009}
\Name{Hinczewski M., Schlagberger X., Rubinstein M., Krichevsky O. \and Netz
  R.~R.} \REVIEW{Macromolecules }{42}{2009}{860}.

\bibitem{Rotne1969}
\Name{Rotne J. \and Prager S.} \REVIEW{J. Chem. Phys. }{50}{1969}{4831}.

\bibitem{Ermak1978}
\Name{Ermak D.~L. \and McCammon J.~A.} \REVIEW{J. Chem. Phys.
  }{69}{1978}{1352}.

\bibitem{Hinczewski2009b}
\Name{Hinczewski M. \and Netz R.~R.} arXiv:0908.0376 (2009).

\bibitem{LeGoff2002}
\Name{Goff L.~L., Hallatschek O., Frey E. \and Amblard F.} \REVIEW{Phys Rev
  Lett }{89}{2002}{258101}.

\bibitem{Harnau1995}
\Name{Harnau L., Winkler R.~G. \and Reineker P.} \REVIEW{J. Chem. Phys.
  }{102}{1995}{7750}.

\bibitem{Wiggins2006}
\Name{Wiggins P.~A., der Heijden T.~V., Moreno-Herrero F., Spakowitz A.,
  Phillips R., Widom J., Dekker C. \and Nelson P.~C.} \REVIEW{Nat. Nanotechnol.
  }{1}{2006}{137}.

\end{thebibliography}

\end{document}